\documentclass[a4paper]{article}

\usepackage{INTERSPEECH2020}
\usepackage{amsmath,graphicx}
\usepackage{xcolor,colortbl}
\usepackage{todonotes}
\usepackage{cite}
\definecolor{Gray}{gray}{0.95}

\newcommand{\xvi}[1]{\mathbf{x}_{#1}}
\newcommand{\zvi}[1]{\mathbf{z}_{#1}}

\newcommand{\zvSi}[1]{\underline{\mathbf{z}}_{#1}}
\newcommand{\xvSRi}[1]{\underline{\hat{\mathbf{x}}}_{#1}}
\newcommand{\xvRi}[1]{\hat{\mathbf{x}}_{#1}}
\newcommand{\Xm}[1]{\mathbf{X}_{#1}}
\newcommand{\Zm}[1]{\mathbf{Z}_{#1}}
\newcommand{\Am}[1]{\mathbf{A}_{#1}}
\newcommand{\Bm}[1]{\mathbf{B}_{#1}}

\title{Self-Expressing Autoencoders for Unsupervised Spoken Term Discovery}
\name{Saurabhchand Bhati$^1$, Jes\'us Villalba$^{1,2}$, Piotr \.Zelasko$^1$, Najim Dehak$^{1,2}$}
\address{$^1$Center for Language and Speech Processing,
 $^2$Human Language Technology Center of Excellence,  \\ Johns Hopkins University, Baltimore, MD, USA}
\email{\{sbhati1,jvillalba,pzelasko,ndehak3\}@jhu.edu}

\begin{document}

\maketitle
\begin{abstract}
Unsupervised spoken term discovery consists of two tasks: finding the acoustic segment boundaries and labeling acoustically similar segments with the same labels. We perform segmentation based on the assumption that the frame feature vectors are more similar within a segment than across the segments. Therefore, for strong segmentation performance, it is crucial that the features represent the phonetic properties of a frame more than other factors of variability. 
We achieve this via a self-expressing autoencoder framework. It consists of a single encoder and two decoders with shared weights. The encoder projects the input features into a latent representation. One of the decoders tries to reconstruct the input from these latent representations and the other from the self-expressed version of them. We use the obtained features to segment and cluster the speech data. We evaluate the performance of the proposed method in the Zero Resource 2020 challenge unit discovery task. The proposed system consistently outperforms the baseline, demonstrating the usefulness of the method in learning representations.
\end{abstract}

\section{Introduction}
Speech recognition technologies have significantly improved in recent years~\cite{synnaeve2019end,wang2019transformer}. Most of the success can be attributed to advancements in deep learning and improved computational resources. However, most state-of-the-art systems require abundant resources including thousands of hours of manually transcribed data~\cite{hannun2014deep,godfrey1992switchboard}, clearly defined lexicon and pronunciation dictionary~\cite{stolcke2011srilm} and substantial amounts of text data for language modeling. As a consequence, the fruits of voice-enabled interfaces, for man-machine interaction, have been limited to resource-rich languages such as English, or Mandarin. The recent technological advances in ASR systems can not be applied to under-resourced languages, for example, regional languages, for which transcribed speech data,  pronunciation dictionary and language models are not readily available.  

Hence, there is a growing interest in developing alternate methods for developing speech processing techniques for low resource languages. Zero Resource speech processing has attracted significant attention in the past several years and aims to develop unsupervised learning techniques that can directly learn from the data~\cite{versteegh2015zero,dunbar2017zero,dunbar2019zero}. It has several applications including the preservation of endangered languages, building speech interfaces in low-resource languages, and developing predictive models for understanding language evolution. 

One of the core issues in zero resource speech processing is the unsupervised discovery of linguistic units from raw speech data~\cite{jansen2011efficient,badino2014auto,huijbregts2011unsupervised,lee2012nonparametric,siu2014unsupervised,kamper2017segmental,bhati2017unsupervised,kamper2017embedded,bhati2018phoneme}. Unit discovery, in turn, requires the waveform to be segmented into acoustically homogeneous regions, consistently labeled based on similar acoustic properties. The representation of the speech signals plays a crucial role in the segmentation and clustering of the speech signal and, in turn, affects the entire process of discovering linguistic units from the speech signal.

The speech waveform carries information about several latent variables, including the identity of the linguistic units, speaker, language, dialect, and emotion, as well as the communication channel and background environment. Consequently, the conventional features extracted from the magnitude spectral envelope of the speech signal capture all these sources.  For example, Mel frequency cepstral coefficients (MFCC) have been successfully used as the speech representation in speech recognition systems~\cite{hinton2012deep}, language identification, emotion recognition~\cite{kwon2003emotion} and speaker-recognition systems~\cite{dehak2010front}, proving their abundance in information about all these tasks. In the supervised scenario, such as supervised ASR, a powerful classifier such as a deep neural network (DNN) is used to learn a non-linear mapping between the input MFCC features and the manual transcripts. Under the supervision of manually marked labels, the classifier can learn to project the MFCC features to an intermediate space, which is speaker-independent. Hence an ASR system can recognize the linguistic units from MFCC features even in the presence of other sources of information~\cite{hinton2012deep}.

However, in an unsupervised scenario, we do not have manual transcriptions to guide us in selecting only the relevant features. Several unsupervised methods have been proposed to learn a representation that highlights the information about the linguistic units by marginalizing the other sources' information~\cite{bhati2018phoneme,zeghidour2016deep,kamper2015unsupervised,thiolliere2015hybrid}. In this paper, we learn embeddings that highlight the speech-specific properties that can help applications like unsupervised spoken term discovery. 

Some side information, like speech segments labeled as the same or different, has been used for representation learning. Given preliminary segments and labels, correspondence autoencoder (CAE)~\cite{kamper2015unsupervised} and ABnet~\cite{thiolliere2015hybrid}  learn feature representations that minimize distances among different instances of the same label. Sia-seg~\cite{bhati2019unsupervised} relaxed the side information requirement and proposed to learn features from just initial segmentation. The features perform better than the raw spectral features at phone discrimination tasks. The quality of the representation depends upon the quality of the side information.  

Auto-encoders have been very successful in unsupervised learning~\cite{badino2015discovering,renshaw2015comparison}. Self reconstruction based losses guide the network to capture the underlying semantic information in the form of compact embeddings. These embeddings capture the underlying phone class along with the background information (speaker, channel, etc.) needed to reconstruct the input as closely as possible. There is no explicit constraint to highlight the underlying phone information. We propose a self-supervised feature learning method that does not require any side information. We call it self-expressing autoencoders (SEA) - a variant of autoencoders that can learn embeddings to capture the underlying phone class information. Our goal is to enforce embeddings for data points belonging to different phone classes to be different. 
SEA consists of an encoder and two decoders with shared weights Fig ~\ref{fig:SAE_arch}. The encoder maps the input features into latent representations. One decoder tries to reconstruct the input from direct latent representations and the other from their self-expressed version. We project the encoder output into two separate latent spaces: one tries to capture the label information, and the other captures the non-label variability. We only impose the self-expression constraint on the label latent space. We expect this leads to the factorization of the label and non-label information. We benchmark the proposed algorithm on the Zero speech 2020 challenge. We achieve excellent performance on the segmentation and clustering tasks without any side information. 

\section{Self-Expressing Autoencoder (SEA)}
Let \(\Xm{}(\in \mathbb{R}^{N\times d}) = \{\xvi{1},\xvi{2},...,\xvi{N}\}\) be a sequence of consecutive speech features where \(\xvi{i}\) is a \(d\) dimensional vector and \(N\) is the total number of frames. These features are generated by a sequence of non-overlapping continuous phone segments \(S = \left\{s_1,s_2,...,s_K\right\}\). The phones can be thought of as the underlying hidden variables that generate the MFCC features. 
Our goal is to train a feature extractor whose latent representation is well suited for segmentation and clustering. A robust representation would yield high similarity between frames representing the same phone, and low similarity for frames representing different phones. In fact, the perfect frame representations would be orthogonal when the underlying phones are different.   

\begin{figure}
    \centering
    \includegraphics[width=\columnwidth]{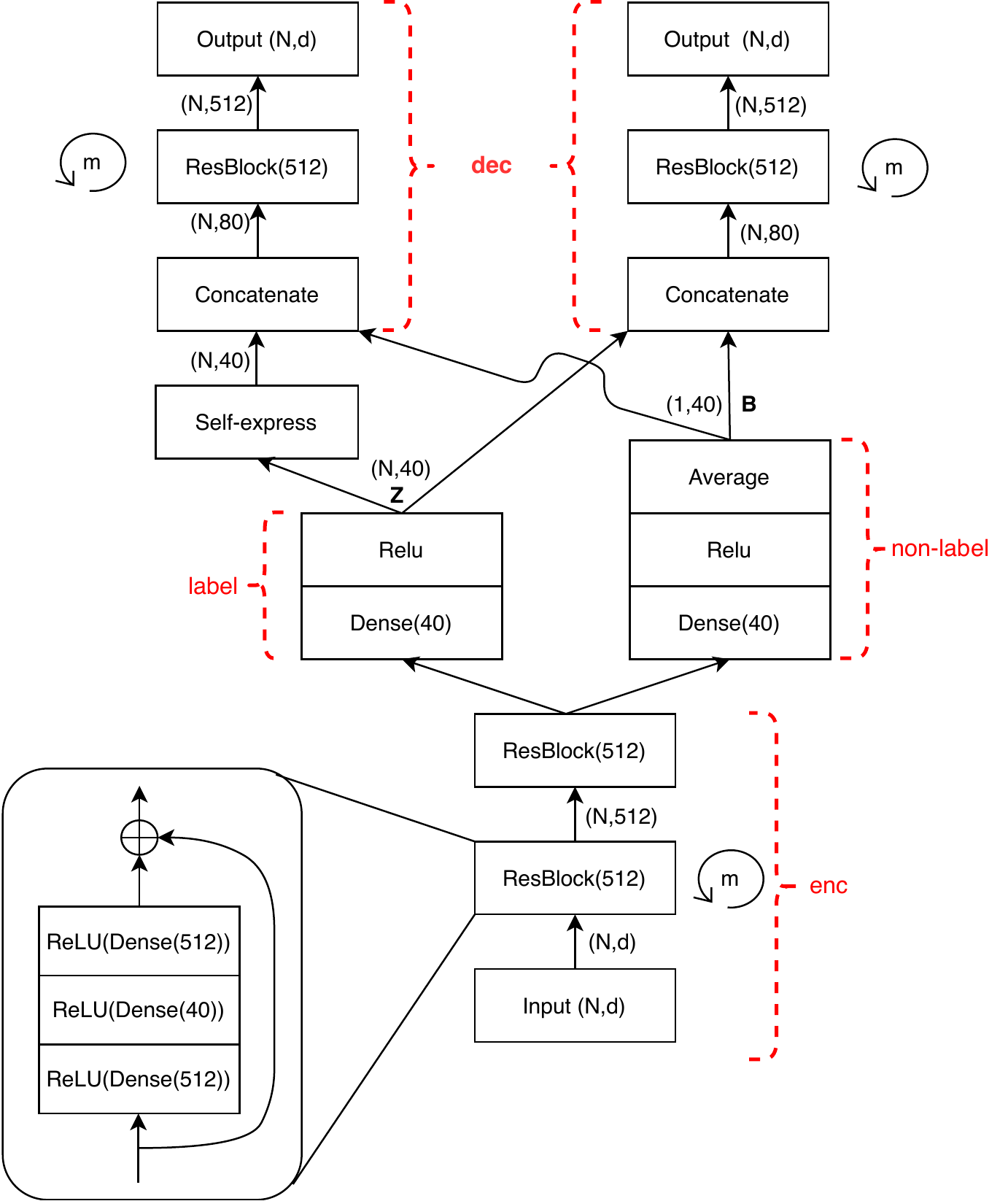}
    \caption{Self-expressing Autoencoder architecture. The parenthesis along with the arrow denotes the output size.}
    \label{fig:SAE_arch}
    \vspace{-13mm}
\end{figure}

Consider an autoencoder architecture with a single encoder and two decoders with tied weights (Fig ~\ref{fig:SAE_arch}). The encoder maps input feature vector, $\xvi{i}$, (e.g. MFCC feature) to embeddings $\zvi{i}$. One decoder takes the direct embedding $\zvi{i}$ to generate reconstruction $\xvRi{i}$ of input $\xvi{i}$ and the other decoder takes the self-expressed version of embedding $\zvSi{i}$ to generate reconstruction $\xvSRi{i}$ of input $\xvi{i}$. The autoencoder tries to make the reconstructions $\xvRi{i}$, $\xvSRi{i}$ as close as possible to the input $\xvi{i}$.

To better understand the method, let us start by considering an example with four frames and two phones. Let $\xvi{1}, \xvi{2}, \xvi{3},\xvi{4}$ be feature vectors that belong to $s_1, s_1, s_2, s_2$ phone classes and $\zvi{1}, \zvi{2},\zvi{3}$ and $\zvi{4}$ be  their corresponding embeddings.
Our observation is that $\xvi{1}, \xvi{2}$ will be more similar to each other than $\xvi{1}, \xvi{3}$ as the feature $\xvi{1}, \xvi{2}$ are drawn from same distribution whereas $\xvi{1}, \xvi{3}$ are drawn from different phone classes. Similar trends are observed in the embeddings. Our goal is to increase the similarity (ideally equal to 1 for cosine metric) between the embeddings for the same class $\zvi{1}$, $\zvi{2}$, and decrease the similarity (ideally equal to 0) between the embeddings for different classes $\zvi{1}$, $\zvi{3}$. 

One decoder tries to reconstruct the input, $\xvi{1}$, from $\zvi{1}$ whereas the other from the self-expressed version given as $\zvSi{1} =  \zvi{1} *\Am{11} + \zvi{2} *\Am{12} + \zvi{3} *\Am{13} + \zvi{4} *\Am{14}$. $\Am{ij}$ denotes the cosine similarity between features $\zvi{i}$ and $\zvi{j}$. The self-similarity matrix, $\Am{}$ captures the relationship between the feature vectors and can be obtained by taking inner product of length normalized features $\Am{} = <\Zm{},\Zm{}>$ or $\Zm{}*\Zm{}^{T}$. 
$\Am{}$ is a symmetric matrix. 

Since the decoders share weights and the desired output for both the decoders is same i.e. $\xvi{i}$, the autoencoder will try to make $\zvi{1}$ and $\zvSi{1}$ same. The network could have simply made $\Am{} \approx I$ to achieve this. But this would not have helped our goal of making $\zvi{1}$ and $\zvi{2}$ similar. In order to prevent a trivial solution, we set $\Am{ii}$ to zero and remove the contribution of $\Am{ii}$ from $\zvSi{i}$.

\begin{align}
    &\zvSi{1} =  \zvi{2} *\Am{12} + \zvi{3} *\Am{13} + \zvi{4} *\Am{14} \\
    &\zvSi{2} =  \zvi{1} *\Am{12} + \zvi{3} *\Am{23} + \zvi{4} *\Am{24} \\
    &\zvSi{3} =  \zvi{1} *\Am{13} + \zvi{2} *\Am{23} + \zvi{4} *\Am{34} \\
    &\zvSi{4} =  \zvi{1} *\Am{14} + \zvi{2} *\Am{24} + \zvi{3} *\Am{34}
\end{align}
The autoencoder will try to force $\zvi{1}$ and $\zvSi{1}$ towards each other. $\zvSi{1}$ only contains $\zvi{2}$, $\zvi{3}$ and $\zvi{4}$. 
Similarly, the encoder will try to force $\zvi{2}$ and $\zvSi{2}$ towards each other. $\zvSi{2}$ only contains $\zvi{1}$, $\zvi{3}$ and $\zvi{4}$. The network will try to push $\zvi{1}$ and $\zvi{2}$ towards each other, thus increasing $\Am{12}$ (ideally equal to 1) and pull $\zvi{1}$ and $\zvi{3}$ away from each other thus decreasing $\Am{13}$ (ideally equal to 0). So by enforcing a self-expressing constraint, we pull the embeddings from the same underlying phone class closer together, and we push the embeddings from different underlying phone classes farther apart.
The background information like speaker or channel might prevent $\Am{13}$ from going all the way down to 0.
If $\zvi{i}$ only contained phone label information then we could achieve perfect self-expression, $\zvi{i} = \zvSi{i}$. This is the basis for our label, non-label factorization. 

We divide the output of the encoder into two parts: label and non-label (Fig~\ref{fig:SAE_arch}). The non-label vector is simply the average along the time axis of the non-label output. By doing this, we hope to ignore the temporally varying label information and capture the non-label information (speaker, channel, etc.) which stays same across the features. We only enforce the self-expressing constraint on the label representation. Since only the phone classes can be self-expressed and the averaged out non-label representation is insufficient for good reconstruction, the network is forced to factor out the label and non-label information across two branches.

The entire learning process is summarized below:

Let $\Xm{N\times d}$ be input spectral features for $N$ continuous frames. label and non-label denote non-linear layer used for dividing the encoder output (Fig.~\ref{fig:SAE_arch}). $\Zm{N\times K}$ is the output of the label and $\Bm{1\times K}$ is the output of the non-label. The last layer has ReLU, non-linear activation, thus making $\Zm{}$ non-negative.   
\begin{align}
    \mathbf{E}_{N\times K} = \mathrm{enc}(\Xm{N\times K}) \\
     \Zm{N\times K} = \mathrm{label}(\mathbf{E}_{N\times K}) \\
    \Bm{1\times K} = \operatorname{non-label}(\mathbf{E}_{N\times K}) 
\end{align}

The matrix $\Am{}$ captures the cosine similarity between different latent embeddings. As $\Zm{}$ is non-negative, then $\Am{}$ will also be non-negative. 
\begin{equation}
    \Am{ij} = \frac{\zvi{i}^{T}\zvi{j}}{\Vert \zvi{i} \Vert \Vert \zvi{j}\Vert}
\end{equation}

Since we do not want a trivial solution, we remove the diagonal entries while self-expressing the latent embeddings. 
\begin{equation}
    \Am{} = \Am{} - \mathbf{I}
\end{equation}
The samples memberships are normalized to sum up to \(1\).

\begin{equation}
    \mathbf{D}_{ii} = \mathrm{diag}(\sum_j \Am{ij})
\end{equation}
\begin{equation}
    \underline{\Zm{}} = (\mathbf{D}^{-1}\Am{})\Zm{}
\end{equation}
In Figure~\ref{fig:SAE_arch}, the two decoder branches share the weights. The direct decoder forces the preservation of all the necessary information required for data reconstruction and the self-expressed branch forces the embeddings to be orthogonal. We concatenate the same non-label vector, $\Bm{}$, to both $\Zm{}$ and $\underline{\Zm{}}$.
\begin{align}
    &\hat{\Xm{}} = \mathrm{dec}(\mathrm{conc}(\Zm{},\Bm{})) \\
    &\hat{\underline{\Xm{}}} = \mathrm{dec}(\mathrm{conc}(\underline{\Zm{}},\Bm{}))
\end{align}

The goal of the network is to force both: the direct reconstruction $\hat{\Xm{}}$ and the self-expressed reconstruction $\hat{\underline{\Xm{}}}$ to be as close as possible to the actual input $\Xm{}$. If the direct embeddings, $\Zm{}$ and their self-expressed version $\underline{\Zm{}}$ are not similar\footnote{\label{note0}
The decoders can map many different inputs to the same output.  
Even though $\Zm{}$ and $\underline{\Zm{}}$ might not match exactly, but their semantic content would match. Our goal is to make the semantic content the same. } then the reconstructions generated from them will be far apart, and the loss will be high. Thus, to minimize the loss, the network has to learn features that are both sufficient for data reconstruction and are orthogonal for various underlying sub-spaces. The network's objective is to minimize the sum of the mean square error between the reconstructions and the input. 

\begin{equation}
    \mathrm{Loss} = \Vert \Xm{} - \hat{\underline{\Xm{}}} \Vert + \Vert \Xm{} - \hat{\Xm{}} \Vert  
    \label{eq:loss}
\end{equation}

The training procedure can also affect the quality of the learned embeddings. We use a two-stage training approach where an autoencoder (one encoder-one decoder) is trained first to minimize a reconstruction loss. We use these weights to initialize SAE and adapt the network. After feature learning, the next step is segmentation into phone-like units.

\section{Segmentation}
Segmentation of the speech signal requires us to detect the time steps at which the vocal tract transitions between states. However, the transition is often smooth rather than abrupt, which makes the detection difficult. To identify the candidate boundaries, we use the Kernel Gram segmentation method~\cite{bhati2018unsupervised}. The algorithm detects the segment boundaries based on the assumption that the frames within a segment exhibit a higher degree of similarity than the frames across the segment. The similarity between two feature vectors $\zvi{i}$ and $\zvi{j}$ is given by:

\begin{equation}
 \mathbf{G}(i,j) = \frac{\zvi{i}^{T}\zvi{j}}{\Vert \zvi{i} \Vert \Vert \zvi{j}\Vert}
\end{equation}

\begin{figure}
    \centering
    \includegraphics[width=\columnwidth]{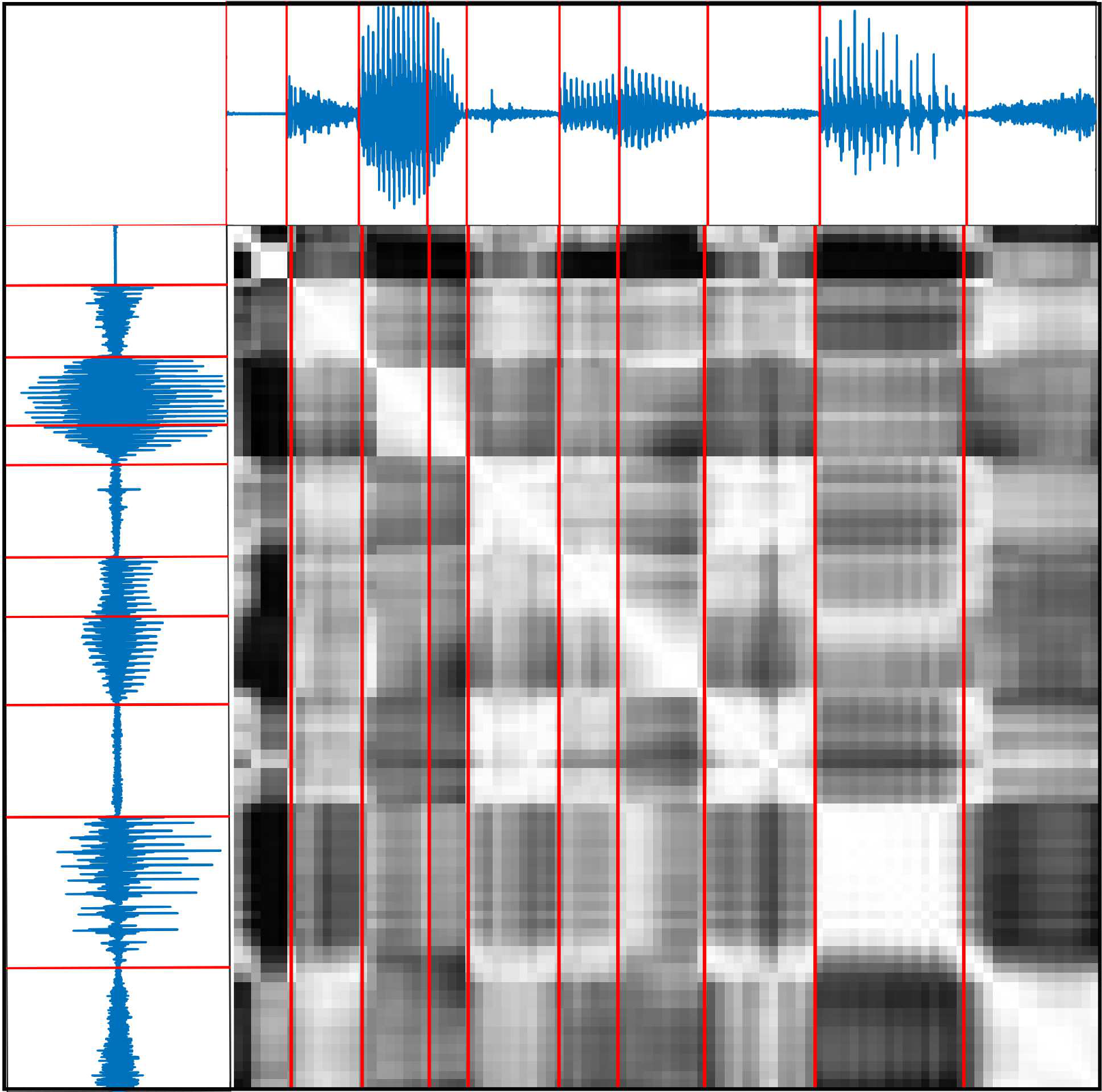}
    \caption{Self-similarity matrix computed from the SAE embeddings. The red lines indicate the manual phoneme boundaries.}
    \label{fig:SS}
    \vspace{-5mm}
\end{figure}

The features belonging to the same segment are acoustically similar and bring about a block diagonal structure in the self-similarity matrix. The task of speech segmentation can now be viewed as identifying the square patches in the Kernel Gram matrix, Fig~\ref{fig:SS}.

The algorithm hypothesizes a frame as a candidate for the boundary if the next $\tau$ consecutive frames have lower similarity than an adaptive threshold $\varepsilon$. A smaller value of $\tau$ would result in false alarms, and a larger $\tau$ would result in missed boundaries. The choice of \(\varepsilon\) affects the boundary detection performance. We use a value of \(\tau = 2\) for boundary detection. The minimum and maximum possible acoustic segment lengths are restricted to 20 ms and 500 ms, respectively.

For all the frames, a prospective boundary is predicted, and frames belonging to the same segment would predict the same or nearby frames as endpoints. For each frame, we count the number of frames that predicted it as the segment endpoint. The frames with a higher count than their adjacent frames are determined as endpoints.
Depending upon the segment in consideration, the level of similarity might vary, e.g., frames within voiced segments are more similar than the unvoiced segments.  An adaptive \(\varepsilon\) that adjusts automatically according to the segment's acoustic properties is required. The \(\varepsilon\) for $i^{th}$ row is set to be the mean of the $i^{th}$ row of the similarity matrix, $\mathbf{G}$. 

\begin{table*}[ht!]
    \centering
    \begin{tabular}{|c|c|c|c|c|c|c|c|c|c|c|c|c|c|c|c|}
    \hline
    Language & System & 
    \multicolumn{2}{c}{NLP} &
    \multicolumn{3}{|c}{type} & 
    \multicolumn{3}{|c}{token} & 
    \multicolumn{3}{|c|}{boundary}\\ 
    \hline 
    & & NED & Cov & P & R & F & P & R & F & P & R & F\\ 
    \hline 
    \hline 
    English & Baseline~\cite{jansen2011efficient} & \textbf{32.42} & 7.93 & \textbf{1.89} & 1.69 & 1.78 & 1.88 & 0.14 & 0.26 & \textbf{32.07} & 3.23 & 5.87 \\ \cline{2-13}
    & SEA &89.5 &\textbf{99.45} &1.73 &\textbf{19.76} &\textbf{3.18} &\textbf{4.44} &\textbf{12.82} &\textbf{6.59} &27.28 &\textbf{75.87} &\textbf{40.13} \\ \hline
    \hline 
    \rowcolor{Gray}
    French & Baseline~\cite{jansen2011efficient} & \textbf{69.5} & 1.59 & \textbf{2.95} & 0.27 & 0.49 & {1.33} & 0.03 & 0.07 & \textbf{32.54} & 0.62 & 1.22 \\ \cline{2-13}
    \rowcolor{Gray}
    & SEA &88.95 & \textbf{99.75}& 2.08 & \textbf{12.93} & \textbf{3.58} & \textbf{4.12} & \textbf{12.98} &\textbf{6.26}& 28.27 & \textbf{81.11} & \textbf{41.93} \\ \hline
    \hline 
    Mandarin & Baseline~\cite{jansen2011efficient} & \textbf{28.6} & 2.71 & 4.87 & 0.15 & 0.28 & 6.37 & 0.1 & 0.2 & \textbf{54.28} & 1.28 & 2.49 \\ \cline{2-13}
    & SEA & 94.64 & \textbf{99.94} &\textbf{6.89} & \textbf{29.08} & \textbf{11.14} & \textbf{7.94} & \textbf{25.43} & \textbf{12.1} & {36.49} & \textbf{91.88} & \textbf{52.23}      
    \\ \hline
    \hline 
    \rowcolor{Gray}    
    LANG1 & Baseline~\cite{jansen2011efficient} & \textbf{32.08} & 3.15 & \textbf{2.86} &  0.52 & 0.88 & 2.84    & 0.09 & 0.17 & \textbf{28.76} & 1.33 & 2.54 \\ \cline{2-13}
    \rowcolor{Gray}
    & SEA & 89.64 & \textbf{99.82} & 1.53 & \textbf{10.01} & \textbf{2.65} & \textbf{4.02} & \textbf{14.58} & \textbf{6.3} & 23.36 & \textbf{80.71} & \textbf{36.24} \\     \hline
    \hline
    LANG 2 & Baseline~\cite{jansen2011efficient} & \textbf{29.57} &    3.39&        4.35    &0.44&    0.8    &4.9&    0.15&    0.29&    \textbf{45.78}&    1.48&    2.87\\ \cline{2-13}
    &  SEA & 82.66 & \textbf{99.85} & \textbf{5.94} & \textbf{34.69} & \textbf{10.14} & \textbf{8.48} & \textbf{24.64} & \textbf{12.61} & {35.65} & \textbf{86.96} & \textbf{50.57} \\ \hline     
    \end{tabular}
    \caption{Performance of the baseline system and proposed SAE embeddings on Zero Resource Speech 2020 Challenge}
    \label{tab:my_label}
    \vspace{-5mm}
\end{table*}

\section{Segment clustering and word discovery}
The segmentation step divides the speech data into a large number of varying length segments. The next step is to cluster the segments into acoustically similar clusters.

\textbf{i) Clustering varying length segments:} To quantify the similarity between two varying length segments, we need to extract fixed-dimensional representation from the varying length segments. Here, we simply took the average of all the features from a segment to obtain a fixed dimensional embedding. We used a Graph-growing algorithm to cluster the segments~\cite{bhati2017unsupervised}. We refer to the clustered segments as virtual phones since they are analogous to actual phones but are discovered by a machine.

\textbf{ii) Word discovery:}
The end goal of the Zero speech 2020 challenge is to discover the word-like units, i.e., ``recurring speech fragments'' from the untranscribed speech data. In our framework, words would correspond to repeating sequences of virtual phones. A bottom-up greedy mapping was used for discovering the words from sequences of virtual phones. First, we find all the longest recurring n-grams that occur at least twice in the data. Then, we locate the n-grams of the next highest order, and virtual phones that were part of the already discovered n-grams are excluded from this step. This process is repeated all the way down to unigrams, and all the remaining unigrams are included as words. The greedy nature of the algorithm makes it computationally efficient, making word discovery feasible even for large datasets. We empirically use tri-grams as the possible words. A similar approach has been previously used for word discovery with success~\cite{bhati2019unsupervised,rasanen2015unsupervised}. 

\section{Experiments}
We used the Zero Resource speech 2020 challenge \footnote{\label{note1}https://zerospeech.com/2020/index.html}for evaluating the proposed approach. The challenge aims to measure the robustness of unsupervised term discovery systems across speakers and languages. The 2020 challenge dataset consisted of more than 100 hours of data distributed across five languages. The vast amount of data ensures that the term discovery systems are scalable to large speech corpora. Three languages (English, French, and Mandarin) were released along with the term discovery evaluation system, for each of them. The system hyper-parameters should be optimized such that the systems generalize well across languages. 

The evaluation kit uses various well-established metrics to quantify the system performance~\cite{ludusan2014bridging}. All the metrics
assume the availability of a time-aligned transcription of the speech data. Normalized edit distance (NED) measures the differences in the phoneme sequences of a word class, while the coverage (Cov) measures the fraction of the data covered by the discovered word-like units.  Most of the measures are defined in terms of precision (P), recall (R), and F-score. Precision (P) is the probability that an element in a discovered set of entities belongs to the gold set, and recall (R) the probability that a gold entity belongs to the discovered set. The F-score is the harmonic mean of precision and recall. The segmentation measures the quality of the boundaries of the identified word-like units with the manual word boundaries.  

Following the guidelines of the challenge, the phonetic segmentation, clustering was done in a speaker and language-independent manner, and the same system (same minimum phoneme length, the same number of clusters, etc.) was used for different languages without optimizing any language-specific parameters. As we can see in Table~\ref{tab:my_label}, the proposed approach consistently achieved better type, token, and boundary F-score on all the languages than the baseline system. Our proposed system focuses on full coverage segmentation and clustering, where entire data is segmented and labeled. The full-coverage system would develop speech indexing and query-by-example~\cite{zhang2009unsupervised} search system in an unsupervised manner.

The baseline system~\cite{jansen2011efficient} used Locality Sensitive Hashing to generate low dimensional bit signatures and then employed segmental DTW for locating matching patterns. The baseline system focuses on finding high-precision isolated segments. To achieve high precision, a lot of the discovered segments are discarded, which results in limited coverage of around 4$\%$. As a result, the baseline system performs better in terms of NED, which is computed only on the discovered patterns. On the other hand, the proposed system covers the entire data and therefore achieves better performance in word boundary, token, type, and coverage. Although the acoustic segments clustered together by our approach had a poorer phonetic match to the ground truth (high NED) as compared to the baseline, the discovered word-like units better matched the true words (word type F-score). We significantly outperformed the baseline algorithm. Our system not only performed well on the training languages but also on the surprise languages, which shows that our system generalizes well across languages and can be used on unseen languages.

\section{Conclusions and Future Work}
Here, we proposed the Self-Expressing autoencoder (SEA) for unsupervised feature learning. We used self-expression as an explicit constraint to highlight the underlying phone class information. 
We successfully applied the learned embeddings for the term discovery task on Zero Resource 2020 challenge. The word discovery pipeline was based on simple heuristics, i.e., all trigrams were considered potential word units. We would like to change it for more advanced methods like ESK-Means~\cite{kamper2017embedded}. In the future, we will experiment with weighting the self-expressed reconstruction loss in the loss term in eq. \ref{eq:loss}. It would also be interesting to analyze the information captured by the non-label branch in the encoder.

\bibliographystyle{ieeetr}
\bibliography{ref}

\begin{thebibliography}{10}

\bibitem{synnaeve2019end}
G.~Synnaeve, Q.~Xu, J.~Kahn, E.~Grave, T.~Likhomanenko, V.~Pratap, A.~Sriram,
  V.~Liptchinsky, and R.~Collobert, ``End-to-end asr: from supervised to
  semi-supervised learning with modern architectures,'' {\em arXiv preprint
  arXiv:1911.08460}, 2019.

\bibitem{wang2019transformer}
Y.~Wang, A.~Mohamed, D.~Le, C.~Liu, A.~Xiao, J.~Mahadeokar, H.~Huang,
  A.~Tjandra, X.~Zhang, F.~Zhang, {\em et~al.}, ``Transformer-based acoustic
  modeling for hybrid speech recognition,'' {\em arXiv preprint
  arXiv:1910.09799}, 2019.

\bibitem{hannun2014deep}
A.~Hannun, C.~Case, J.~Casper, B.~Catanzaro, G.~Diamos, E.~Elsen, R.~Prenger,
  S.~Satheesh, S.~Sengupta, A.~Coates, {\em et~al.}, ``Deep speech: Scaling up
  end-to-end speech recognition,'' {\em arXiv preprint arXiv:1412.5567}, 2014.

\bibitem{godfrey1992switchboard}
J.~J. Godfrey, E.~C. Holliman, and J.~McDaniel, ``Switchboard: Telephone speech
  corpus for research and development,'' in {\em Acoustics, Speech, and Signal
  Processing, 1992. ICASSP-92., 1992 IEEE International Conference on}, vol.~1,
  pp.~517--520, IEEE, 1992.

\bibitem{stolcke2011srilm}
A.~Stolcke, J.~Zheng, W.~Wang, and V.~Abrash, ``Srilm at sixteen: Update and
  outlook,'' in {\em Proceedings of IEEE Automatic Speech Recognition and
  Understanding Workshop}, vol.~5, 2011.

\bibitem{versteegh2015zero}
M.~Versteegh, R.~Thiolliere, T.~Schatz, X.-N. Cao, X.~Anguera, A.~Jansen, and
  E.~Dupoux, ``The zero resource speech challenge 2015.,'' in {\em
  INTERSPEECH}, pp.~3169--3173, 2015.

\bibitem{dunbar2017zero}
E.~Dunbar, X.~N. Cao, J.~Benjumea, J.~Karadayi, M.~Bernard, L.~Besacier,
  X.~Anguera, and E.~Dupoux, ``The zero resource speech challenge 2017,'' in
  {\em Automatic Speech Recognition and Understanding Workshop (ASRU), 2017
  IEEE}, pp.~323--330, IEEE, 2017.

\bibitem{dunbar2019zero}
E.~Dunbar, R.~Algayres, J.~Karadayi, M.~Bernard, J.~Benjumea, X.-N. Cao,
  L.~Miskic, C.~Dugrain, L.~Ondel, A.~W. Black, {\em et~al.}, ``The zero
  resource speech challenge 2019: Tts without t,'' {\em arXiv preprint
  arXiv:1904.11469}, 2019.

\bibitem{jansen2011efficient}
A.~Jansen and B.~Van~Durme, ``Efficient spoken term discovery using randomized
  algorithms,'' in {\em Automatic Speech Recognition and Understanding (ASRU),
  2011 IEEE Workshop on}, pp.~401--406, IEEE, 2011.

\bibitem{badino2014auto}
L.~Badino, C.~Canevari, L.~Fadiga, and G.~Metta, ``An auto-encoder based
  approach to unsupervised learning of subword units,'' in {\em Acoustics,
  Speech and Signal Processing (ICASSP), 2014 IEEE International Conference
  on}, pp.~7634--7638, IEEE, 2014.

\bibitem{huijbregts2011unsupervised}
M.~Huijbregts, M.~McLaren, and D.~Van~Leeuwen, ``Unsupervised acoustic sub-word
  unit detection for query-by-example spoken term detection,'' in {\em
  Acoustics, Speech and Signal Processing (ICASSP), 2011 IEEE International
  Conference on}, pp.~4436--4439, IEEE, 2011.

\bibitem{lee2012nonparametric}
C.-y. Lee and J.~Glass, ``A nonparametric bayesian approach to acoustic model
  discovery,'' in {\em Proceedings of the 50th Annual Meeting of the
  Association for Computational Linguistics: Long Papers-Volume 1}, pp.~40--49,
  Association for Computational Linguistics, 2012.

\bibitem{siu2014unsupervised}
M.-h. Siu, H.~Gish, A.~Chan, W.~Belfield, and S.~Lowe, ``Unsupervised training
  of an hmm-based self-organizing unit recognizer with applications to topic
  classification and keyword discovery,'' {\em Computer Speech \& Language},
  vol.~28, no.~1, pp.~210--223, 2014.

\bibitem{kamper2017segmental}
H.~Kamper, A.~Jansen, and S.~Goldwater, ``A segmental framework for
  fully-unsupervised large-vocabulary speech recognition,'' {\em Computer
  Speech \& Language}, vol.~46, pp.~154--174, 2017.

\bibitem{bhati2017unsupervised}
S.~Bhati, S.~Nayak, and K.~S.~R. Murty, ``Unsupervised speech signal to symbol
  transformation for zero resource speech applications,'' {\em Proc.
  Interspeech 2017}, pp.~2133--2137, 2017.

\bibitem{kamper2017embedded}
H.~Kamper, K.~Livescu, and S.~Goldwater, ``An embedded segmental k-means model
  for unsupervised segmentation and clustering of speech,'' in {\em 2017 IEEE
  Automatic Speech Recognition and Understanding Workshop (ASRU)},
  pp.~719--726, IEEE, 2017.

\bibitem{bhati2018phoneme}
S.~Bhati, H.~Kamper, and K.~S.~R. Murty, ``Phoneme based embedded segmental
  k-means for unsupervised term discovery,'' in {\em 2018 IEEE International
  Conference on Acoustics, Speech and Signal Processing (ICASSP)},
  pp.~5169--5173, IEEE, 2018.

\bibitem{hinton2012deep}
G.~Hinton, L.~Deng, D.~Yu, G.~E. Dahl, A.-r. Mohamed, N.~Jaitly, A.~Senior,
  V.~Vanhoucke, P.~Nguyen, T.~N. Sainath, {\em et~al.}, ``Deep neural networks
  for acoustic modeling in speech recognition: The shared views of four
  research groups,'' {\em IEEE Signal Processing Magazine}, vol.~29, no.~6,
  pp.~82--97, 2012.

\bibitem{kwon2003emotion}
O.-W. Kwon, K.~Chan, J.~Hao, and T.-W. Lee, ``Emotion recognition by speech
  signals,'' in {\em Eighth European Conference on Speech Communication and
  Technology}, 2003.

\bibitem{dehak2010front}
N.~Dehak, P.~J. Kenny, R.~Dehak, P.~Dumouchel, and P.~Ouellet, ``Front-end
  factor analysis for speaker verification,'' {\em IEEE Transactions on Audio,
  Speech, and Language Processing}, vol.~19, no.~4, pp.~788--798, 2010.

\bibitem{zeghidour2016deep}
N.~Zeghidour, G.~Synnaeve, M.~Versteegh, and E.~Dupoux, ``A deep scattering
  spectrum—deep siamese network pipeline for unsupervised acoustic
  modeling,'' in {\em Acoustics, Speech and Signal Processing (ICASSP), 2016
  IEEE International Conference on}, pp.~4965--4969, IEEE, 2016.

\bibitem{kamper2015unsupervised}
H.~Kamper, M.~Elsner, A.~Jansen, and S.~Goldwater, ``Unsupervised neural
  network based feature extraction using weak top-down constraints,'' in {\em
  Acoustics, Speech and Signal Processing (ICASSP), 2015 IEEE International
  Conference on}, pp.~5818--5822, IEEE, 2015.

\bibitem{thiolliere2015hybrid}
R.~Thiolliere, E.~Dunbar, G.~Synnaeve, M.~Versteegh, and E.~Dupoux, ``A hybrid
  dynamic time warping-deep neural network architecture for unsupervised
  acoustic modeling.,'' in {\em INTERSPEECH}, pp.~3179--3183, 2015.

\bibitem{bhati2019unsupervised}
S.~Bhati, S.~Nayak, K.~S.~R. Murty, and N.~Dehak, ``Unsupervised acoustic
  segmentation and clustering using siamese network embeddings,'' {\em Proc.
  Interspeech 2019}, pp.~2668--2672, 2019.

\bibitem{badino2015discovering}
L.~Badino, A.~Mereta, and L.~Rosasco, ``Discovering discrete subword units with
  binarized autoencoders and hidden-markov-model encoders.,'' in {\em
  INTERSPEECH}, pp.~3174--3178, 2015.

\bibitem{renshaw2015comparison}
D.~Renshaw, H.~Kamper, A.~Jansen, and S.~Goldwater, ``A comparison of neural
  network methods for unsupervised representation learning on the zero resource
  speech challenge.,'' in {\em INTERSPEECH}, pp.~3199--3203, 2015.

\bibitem{bhati2018unsupervised}
S.~Bhati, S.~Nayak, and K.~Sri Rama~Murty, ``Unsupervised segmentation of
  speech signals using kernel-gram matrices,'' in {\em Computer Vision, Pattern
  Recognition, Image Processing, and Graphics: 6th National Conference,
  NCVPRIPG 2017, Mandi, India, December 16-19, 2017, Revised Selected Papers
  6}, pp.~139--149, Springer, 2018.

\bibitem{rasanen2015unsupervised}
O.~R{\"a}s{\"a}nen, G.~Doyle, and M.~C. Frank, ``Unsupervised word discovery
  from speech using automatic segmentation into syllable-like units.,'' in {\em
  INTERSPEECH}, pp.~3204--3208, 2015.

\bibitem{ludusan2014bridging}
B.~Ludusan, M.~Versteegh, A.~Jansen, G.~Gravier, X.-N. Cao, M.~Johnson, and
  E.~Dupoux, ``Bridging the gap between speech technology and natural language
  processing: an evaluation toolbox for term discovery systems,'' in {\em
  Language Resources and Evaluation Conference}, 2014.

\bibitem{zhang2009unsupervised}
Y.~Zhang and J.~R. Glass, ``Unsupervised spoken keyword spotting via segmental
  dtw on gaussian posteriorgrams,'' in {\em Automatic Speech Recognition \&
  Understanding, 2009. ASRU 2009. IEEE Workshop on}, pp.~398--403, IEEE, 2009.

\end{thebibliography}

\end{document}